\documentclass[aps,prl,10pt,twocolumn,showpacs,preprintnumbers,amsmath,amssymb,superscriptaddress]{revtex4-1}
\usepackage[english]{babel}
\usepackage[dvips]{graphicx}
\usepackage{dcolumn}
\usepackage{bm}

\usepackage{color}

\newcommand{\R}[1]{\textcolor{black}{#1}}

\begin{document}
\title{Experimental observation of a current-driven instability \\ in a neutral electron-positron beam}
\author{J. Warwick}
\affiliation{School of Mathematics and Physics, Queen's University Belfast, University Road,
Belfast BT7 1NN, UK}
\author{T. Dzelzainis}
\affiliation{School of Mathematics and Physics, Queen's University Belfast, University Road,
Belfast BT7 1NN, UK}
\author{M .E. Dieckmann}
\affiliation{Department of Science and Technology (ITN), Link\"oping University, Campus Norrk\"oping, 60174 Norrk\"oping, Sweden}
\author{W. Schumaker}
\affiliation{SLAC National Accelerator Laboratory, Menlo Park, California 94025, USA}
\author{D. Doria}
\affiliation{School of Mathematics and Physics, Queen's University Belfast, University Road,
Belfast BT7 1NN, UK}
\author{L. Romagnani}
\affiliation{LULI, Ecole Polytechnique, CNRS, CEA, UPMC, 91128 Palaiseau, France}
\author{K. Poder}
\affiliation{The John Adams Institute for Accelerator Science, Blackett Laboratory, Imperial
College London, London SW72BZ, UK}
\author{J. M. Cole}
\affiliation{The John Adams Institute for Accelerator Science, Blackett Laboratory, Imperial
College London, London SW72BZ, UK}
\author{A. Alejo}
\affiliation{School of Mathematics and Physics, Queen's University Belfast, University Road,
Belfast BT7 1NN, UK}
\author{M. Yeung}
\affiliation{School of Mathematics and Physics, Queen's University Belfast, University Road,
Belfast BT7 1NN, UK}
\author{K.Krushelnick}
\affiliation{Center for Ultrafast Optical Science, University of Michigan, Ann Arbor, Michigan
481099-2099, USA}
\author{S.P.D. Mangles}
\affiliation{The John Adams Institute for Accelerator Science, Blackett Laboratory, Imperial
College London, London SW72BZ, UK}
\author{Z. Najmudin}
\affiliation{The John Adams Institute for Accelerator Science, Blackett Laboratory, Imperial
College London, London SW72BZ, UK}
\author{B. Reville}
\affiliation{School of Mathematics and Physics, Queen's University Belfast, University Road,
Belfast BT7 1NN, UK}
\author{G. M. Samarin}
\affiliation{School of Mathematics and Physics, Queen's University Belfast, University Road,
Belfast BT7 1NN, UK}
\author{D. Symes}
\affiliation{Central Laser facility, Rutherford Appleton Laboratory, Didcot, Oxfordshire OX11
0QX, UK}
\author{A. G. R. Thomas}
\affiliation{Center for Ultrafast Optical Science, University of Michigan, Ann Arbor, Michigan
481099-2099, USA}
\affiliation{Lancaster University, Lancaster LA1 4YB, United Kingdom}
\author{M. Borghesi}
\affiliation{School of Mathematics and Physics, Queen's University Belfast, University Road,
Belfast BT7 1NN, UK}
\author{G.Sarri$^*$}
\affiliation{School of Mathematics and Physics, Queen's University Belfast, University Road,
Belfast BT7 1NN, UK}
\email{corresponding author}

\date{\today}
\begin{abstract}
We report on the first experimental observation of a current-driven instability developing in a quasi-neutral matter-antimatter beam. Strong magnetic  fields ($\geq$ 1 T) are measured, via means of a proton radiography technique, after the propagation of a neutral electron-positron beam through a background electron-ion plasma. The experimentally determined equipartition parameter of $\epsilon_B \approx 10^{-3}$, is typical of values inferred from models of astrophysical gamma-ray bursts, in which the relativistic flows are also expected to be pair dominated. The data, supported by Particle-In-Cell simulations and simple analytical estimates, indicate that these magnetic fields persist in the background plasma for thousands of inverse plasma frequencies. The existence of such long-lived magnetic fields can be related to analog astrophysical systems, such as those prevalent in lepton-dominated jets.
\end{abstract}
\maketitle

The exact symmetry between its positively and negatively charged constituents makes electron-positron plasmas and beams (EPBs) unique cases in plasma physics. 
\R{For instance, the exact mobility of the two species forbids the excitation of drift and acoustic modes \cite{comments}, and, more generally, EPBs have a much more simplified Clemmow-Mullaly-Allis diagram than that of their electron-ion counterpart (see, for instance, \cite{Stenson}).} EPBs are also believed to play a central role in a range of high-energy astrophysical phenomena, such as the ultra-relativistic outflows from active galactic nuclei and pulsars \cite{Blandford,Begelman,Goldreich,Wardle}. It has been proposed that pair-dominated jets might play a role in the emission of Gamma-Ray Bursts (GRBs), produced in compact object mergers/collisions, or during the death of massive stars. These events account for some of the most luminous events in the universe \cite{Piran, Racusin}.

Arguably, the most fundamental open question regarding electron-positron beams in astrophysical systems concerns their interaction with the ambient environment, and in particular the related growth of plasma instabilities and magnetic field amplification \cite{Bret,Medvedev,Reville}, essential ingredients in the formation of collisionless shocks and their radiative emission \cite{Lyubarsky,Gruzinov}.  These phenomena require magnetic energy densities with values greatly exceeding that of the ambient plasma (typically with a mean field on the order of a nT \cite{Ferriere}). In these scenarios, the strength of the magnetic fields is usually given in terms of the so-called equipartition parameter $\epsilon_B = U_B/U_e$, with $U_B = B^2/2\mu_0$ and $U_e = \gamma_b n_b m_e c^2$ the magnetic and total kinetic energy density, respectively (here $n_b$ and $\gamma_b$ refer to the density and bulk Lorentz factor of the beam or shock, respectively). GRB afterglow spectra must be the result of synchrotron radiation in a magnetic field with $\epsilon_B$ ranging from $10^{-5}$ \cite{Galama, Vreeswijk} to 0.1 \cite{Waxman,Wijers}. These values cannot be obtained by MHD shock compression of the local magnetic fields ($\epsilon_B \approx 10^{-11}$ \cite{Sari}) or by magnetic flux carried from the central engine ($\epsilon_B  < 10^{-7}$ \cite{Meszaros}). Weibel-mediated shocks in the jet could generate fields of sufficient strength, but are expected to decay rapidly, on timescales comparable to the inverse plasma frequency \cite{Gruzinov}. On the other hand, analytical \cite{Medvedev} and numerical \cite{Chang,Muggli,Silva,Dieckmann} studies give significant evidence that magnetic fields of sufficient strength and persistence might be generated by strong current filamentation of the EPB.

\R{To date, there is no direct evidence of these phenomena, either in the laboratory or in astrophysical observations. This lack of experimental data is ultimately due to the difficulty of generating neutral EPBs in the laboratory despite considerable dedicated efforts worldwide \cite{Chen,Pedersen,Liang}. 
It is only recently that the first generation of a quasi-neutral EPB  \cite{SarriNcomm,SarriPPCF}  in a fully laser-driven setup \cite{SarriPRL} have been reported.}
 
\R{In this Letter, we present the first experimental observation of kinetic behaviour of a neutral pair beam in the laboratory. Strong ($\epsilon_B\approx10^{-3}$) and long-lived (persisting for at least $(2.5\pm0.5)\times10^3$ inverse plasma frequencies of the background plasma) magnetic fields are detected after the propagation of a neutral EPB through a background electron-ion plasma. Analytical considerations and numerical simulations indicate that these fields, observed via a proton imaging technique \cite{SarriNJP,Borghesi}, are the result of a current-driven transverse instability in the beam. The present experiment opens up the possibility of studying phenomena of direct relevance to pair-dominated astrophysical scenarios in the laboratory.}

\begin{figure}[!t]
\begin{center}
\includegraphics[width=1.05\columnwidth]{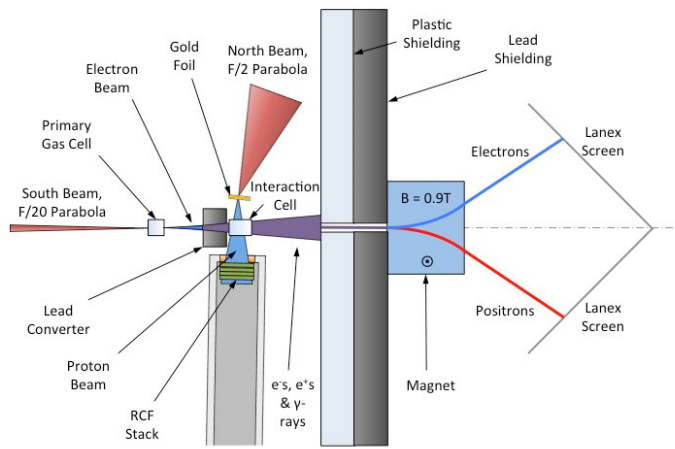}
\caption{Sketch of the experimental setup.}
\label{setup}
\end{center}
\end{figure}

The experiment was carried out using the Astra-Gemini laser \cite{Gemini} hosted by the Central Laser Facility at the Rutherford Appleton Laboratory, UK. The experimental setup is sketched in Fig. \ref{setup}. A short (pulse duration of $45\pm2$ fs) laser pulse, containing an energy of approximately 9 J, was focussed, using an F/20 off-axis parabola, down to a focal spot of diameter $27\pm5$ $\mu$m at the entrance of a 10mm-long gas-cell filled with a He gas doped with 3\% of N$_2$. The helium gas was fully ionized by the laser pulse, producing a plasma density of $4\times10^{18}$ cm$^{-3}$, as measured by optical interferometry. The interaction generated, via laser-wakefield acceleration \cite{Esarey}, a reproducible electron beam with a broad spectrum extending to approximately 600 MeV and an overall charge of the order of $0.40\pm0.04$ nC (similar to what reported in Ref. \cite{SarriNcomm}). \R{The electron beam properties are consistent with results reported in the literature for similar laser parameters, in a regime of ionisation injection (see, for instance, Refs. \cite{Clayton,Kneip}).}
The electron beam then interacted with a lead target with a variable thickness (ranging from 5 to 25mm) in order to generate an EPB that subsequently propagated through a secondary gas-cell filled with pure He. By changing the thickness of the converter target, the percentage of positrons in the EPB can be controlled, seamlessly, from 0\% to approximately 50\% \cite{SarriNcomm,SarriPPCF,SarriPRL}. \R{The rear side of the gas-cell was covered with a LANEX scintillator screen, in order to provide information about the spatial distribution of the EPB with and without the second gas-cell. During free propagation, the EPB presents a smooth spatial profile, well approximated by a super-gaussian \cite{SarriPPCF2}.} Finally, a magnetic spectrometer allowed separating and measuring the spectrum of the electrons and positrons in the beam \R{on each laser shot}. \R{This spectrometer consisted of a 10~cm long, 0.9 T dipole magnet followed by two LANEX stintillators. The scintillators were cross-calibrated using image plates in order to extract the absolute total charge of the electron and positron populations in the EPB.}

A second laser pulse (pulse duration of 45$\pm$2 fs and energy of 9J) was focussed, using an F/2 off-axis parabola, on the surface of a 15 $\mu$m thick gold foil in order to generate, via Target Normal Sheath Acceleration \cite{BorghesiRMP}, a multi-MeV proton beam  with a cut-off energy of $\simeq$ 5 MeV \R{and a smooth spatial profile \cite{suppTIME}}. This beam provided temporally-resolved proton radiographs \cite{SarriNJP} of the plasma, transverse to the EPB propagation, with a geometrical magnification $M\approx8$ \cite{SarriNJP}. 
In this manuscript we focus our attention on radiographs of the same interaction obtained with proton energies of 4.5, 3.3, and 1.1 MeV (each with an uncertainty of 0.5 MeV \cite{SarriNJP}). These energies correspond to probing the background plasma $(14\pm6)$ ps, $(60\pm10)$ ps, and $(280\pm30)$ ps after the transit of the EPB. \R{Monte-Carlo simulations using the code SRIM \cite{SRIM} indicate a broadening of the probing proton beam caused by lateral straggling whilst propagating through the gas-fill in the cell. This broadening is of the order of 2 $\mu$m for a 3.3 MeV proton (5$\mu$m for 1.1 MeV) at the rear side of the gas-cell. This uncertainty is smaller than the intrinsic spatial resolution of the radiographic technique (of the order of 10 $\mu$m for our experimental parameters \cite{SarriNJP}), and it will thus be neglected hereafter.}

For a converter thickness of 2.5 cm (corresponding to approximately 5 radiation lengths), an EPB with $N_e = (3.2\pm0.3)$ $\times10^9$ electrons, $N_p=(3.0\pm0.2)$ $\times10^9$ positrons (positrons accounting for 48$\pm$5\% of the overall leptonic beam) was consistently generated. \R{The electrons and positrons presented a broad spectrum well approximated by a J\"uttner-Synge distribution (average Lorentz factor $\gamma_b \approx 15$), similarly to that reported in Ref. \cite{SarriNcomm}. Matching Monte-Carlo simulations using the code FLUKA \cite{FLUKA} indicate an average divergence of the order of 30 - 50 mrad, and a source size of the order of 300 $\mu$m, whilst analytical estimates of the cascade in the solid \cite{SarriPPCF} indicate a beam duration at source of the order of $\tau_b\approx100$ fs}. The number density of the EPB at the entrance of the second gas-cell, placed 7 mm away from the rear surface of the converter target, is then $n_b$ $= (2.6\pm0.5)\times 10^{14}$ cm$^{-3}$. The EPB co-propagated with an intense burst of bremsstrahlung $\gamma$-rays as predicted by FLUKA simulations \cite{FLUKA}.  Hydrodynamic simulations (using the commercial code HYADES \cite{HYADES}) indicate that this photon beam fully ionises the He gas in the second gas-cell to an average electron density of $n_{\rm pl} = 10^{17}$ cm$^{-3}$ (corresponding electron plasma frequency $\omega_{\rm pl} \approx 2\times 10^{13}$ Hz) and a temperature of the order of 10-20 eV \R{\cite{suppHYADES}}.

\begin{figure}[!t]
\begin{center}
\includegraphics[width=1\columnwidth]{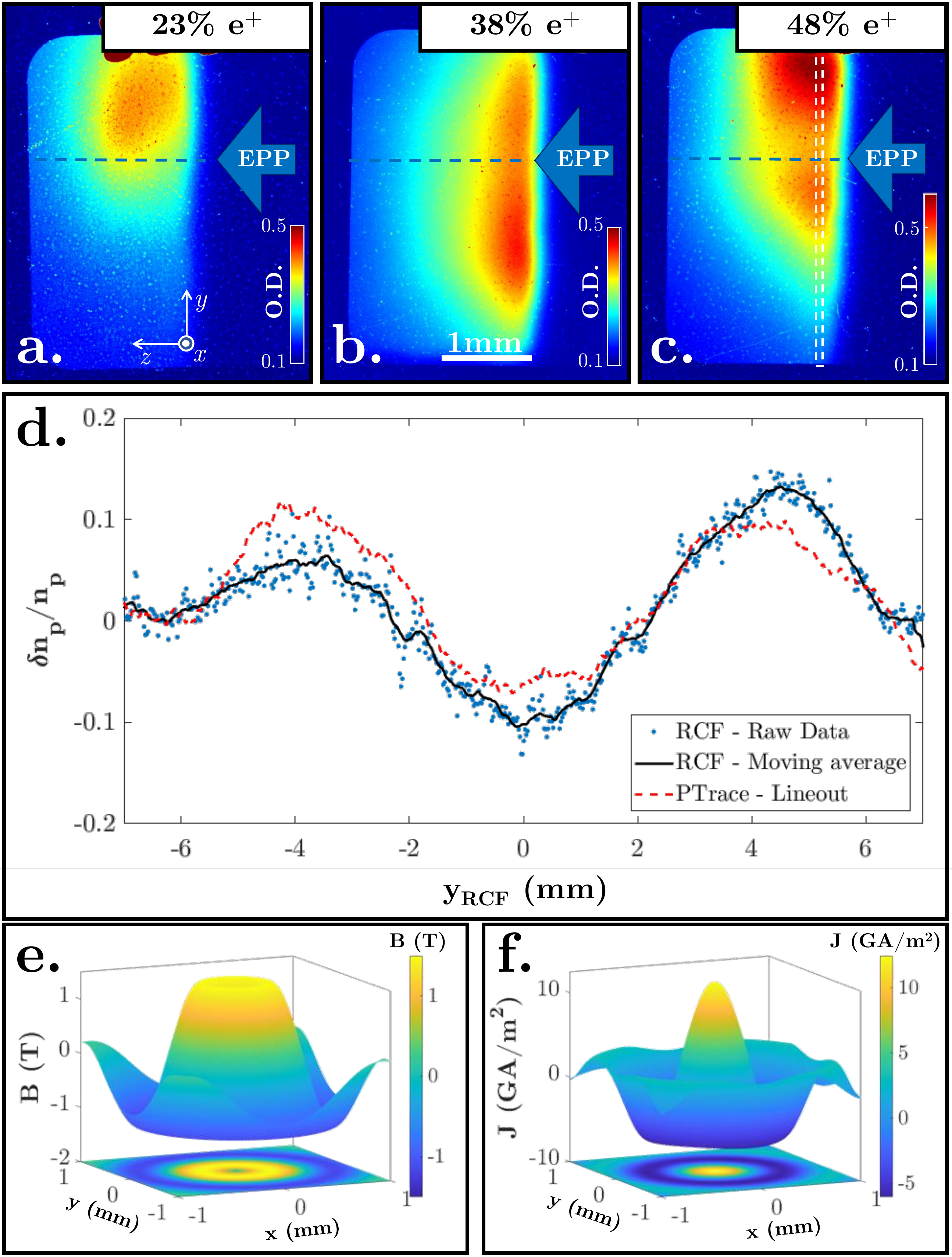}
\caption{\textbf{a.} - \textbf{c.} Typical optical density of the proton radiographies of the background gas after the passage of the electron-positron beam for different percentages of positrons in the beam: 23\% (a.), 38\% (b.), and 48\% (c.). The beam propagates from right to left, as indicated by the arrow, with the main propagation axis represented by the dashed blue line. The spatial scale is common for all frames and refers to the interaction plane. Each radiograph is taken (280$\pm$30) ps after the transit of the EPB (corresponding proton energy of (1.1$\pm$ 0.5) MeV). \textbf{d.} Comparison between the experimental proton distribution and the output of the particle-tracing simulation for frame c. The lineout position is highlighted by the white dashed rectangle in frame c. and it is taken at the detection plane, with the spatial scale thus magnified by a factor M~8. \textbf{e.} Distribution of the azimuthal magnetic field used as an input for the particle-tracing simulation and \textbf{f.} related current density. }
\label{RCFneutral}
\end{center}
\end{figure}

Fig.\ref{RCFneutral} shows typical radiographs of the He plasma (280$\pm$30) ps after the propagation of the EPB. For a low percentage of positrons (frame a.) no proton deflections are observed, with the probing proton beam retaining a smooth spatial profile. As the positron percentage in the EPB is increased (thicker converter target, see Ref. \cite{SarriNcomm}) a faint modulation starts to be observed along the vertical axis (frame b.), which becomes apparent whenever the EPB approaches overall charge neutrality (frame c.).  
Radiographs of the same shots at 3.3 MeV and 4.5 MeV (corresponding to probing times of (60$\pm$10) ps and (14$\pm$6) ps, respectively) show analogous deflection patterns, strong indication of the persistence of the fields responsible for the proton deflections \R{(see Ref. \cite{suppTIME})}. These radiographs are taken long after the EPB has escaped the probed region and we then ascribe the proton deflections to magnetic fields left in the background plasma in the wake of the EPB. \R{Electrostatic fields could also arise from charge separation induced by the EPB in the background plasma, within a typical length-scale $r_E\approx \sigma_b(n_b/n_{pl})^{1/2}\approx$ 25 $\mu$m \cite{Joshi}. This length is comparable to the plasma wavelength ($\ell_{pl}\approx 17$ $\mu$m) indicating that charge neutrality will be restored on a time comparable to the ion plasma period ($\omega_{pi}^{-1}\approx 2$ ps). At the time of observation, we would then expect no significant electrostatic fields left in the plasma. This is confirmed by the experimental data, since hypothetic electrostatic deflections would show for each positron percentage, a feature that is absent in the data (see Fig. \ref{RCFneutral}).}

In order to extract the magnetic field distribution responsible for the observed proton deflections (see Ref. \cite{Smyth} for more details), Particle Tracing (PT) calculations were performed. The best match with the experimental data is obtained for a magnetic field distribution as shown in Fig. \ref{RCFneutral}.e. The field has a peak amplitude of $(1.2\pm0.5)$ T and a characteristic spatial scale of $\lambda_{\rm fil} =1.2\pm0.2$ mm (see Figs. \ref{RCFneutral}.d and \ref{RCFneutral}.e). Within the experimental uncertainty, the same magnetic field distribution reproduces the proton deflections also for proton energies of 3.3 and 4.5 MeV, indicating that the magnetic field does not change significantly over a probing temporal window of $(2.5\pm0.5)\times10^3$ inverse plasma frequencies of the background plasma. The related current density is made of a positive component peaking at $J_{\rm max}\approx 2\times10^{10}$ A/m$^2$ in the center (corresponding to a particle density of $n_J\approx2\times10^{14}$ cm$^{-3}$), surrounded by a negative current (Fig. \ref{RCFneutral}.f). Given the time-scale of the observation, this corresponds to the return currents left in the background plasma after the propagation of the EPB. Since $n_J\approx n_b$, this is consistent with the EPB creating only one large filamentary structure. In our case of a weak beam ($\alpha=n_b/n_{pl}\ll$ 1), this current-driven instability \cite{filamentation} is expected to generate a transverse modulation with a growth rate of the order of $\Gamma_{\rm fil}\approx \omega_{\rm pl}\sqrt{\alpha/\gamma_b}\approx  2\times10^{11}$ Hz (growth time of approximately 5 ps). The instability thus takes only 1.5mm to develop, well within the EPB propagation distance observed in the RCFs. Also the measured spatial scale is comparable to twice the beam skin depth, which is of the order of 600 $\mu$m.

It must be noted that, for a magnetic field of approximately 1.2 T,  the Larmor radius of the background electrons (approximately 30 $\mu$m for a simulated background temperature of 10 eV) is much smaller than the typical spatial scale of the magnetic field (of the order of 1 mm), indicating that the background plasma can effectively get magnetised. 
Moreover, the magnetic field in the background plasma will dissipate only via resistive effects. This is because collisionless dissipation is ruled out, since the spatial scale of the field is much larger than the skin depth of the background plasma (a few microns). In this regime, the temporal scale for magnetic field dissipation can be estimated as $\tau_{\rm OHM}\approx \mu_o\sigma\lambda_{\rm fil}^2$, with $\sigma = n_e e^2/(m_e\nu_{\rm ei})$ the classical conductivity of the plasma. For our parameters, the electron-ion collision frequency is of the order of $\nu_{\rm ei}\approx9\times10^8$ Hz, implying a temporal scale of $\tau_{\rm OHM}\approx$ 75 $\mu$s. This time is much larger than our observation time, justifying why the field is experimentally seen to retain its shape and amplitude.

\R{The interpretation of a purely transverse current-driven instability is also corroborated by the spectral and spatial distribution of the EPB after propagation through the background plasma. Within the experimental uncertainties, no significant modulation is seen in the spectrum of the EPB, indication of no longitudinal modes being excited \cite{Sironi}. This is to be expected since the beam is longitudinally much shorter than its skin depth. Moreover the scintillator screen placed at the exit of the gas-cell does not show any spatial modulation, with the EPB retaining its smooth spatial distribution \cite{SarriPPCF2}. Numerical simulations \cite{SarriNcomm} show that, due to both species in the beam having the same mobility, electron and positron filaments will distribute symmetrically, thus retaining a smooth number density in the beam. In the ultra-relativistic regime, a scintillator screen would not be sensitive to the sign of the charge or the energy of the particle but only on the number of particles, explaining why no density modulations are observed \cite{suppCHARGE}. Indeed, this is further confirmation of observing a purely current-driven instability.}

In order to support the interpretation of the experiment, we performed a 2-dimensional Particle-In-Cell (PIC) simulation using the EPOCH code \cite{Arber15}. The simulation box resolves the intervals 0 $\le x \le $ 10 mm along the beam propagation direction and -1.5 mm $\le y \le $ 1.5 mm orthogonal to it by $10^4$ grid cells and $3 \times 10^3$ grid cells, respectively. We use open boundary conditions for the particles and fields. The pair cloud consists of electrons and positrons with a mean Lorentz factor of $\gamma=15$. The positron density distribution at the time $t_0=0$ is  $n_p(x,y,t_0) = n_0 \exp{(-y^2/c_p^2)}$ with $c_p = 118 \mu$m, and $n_0 = 10^{16}\textrm{cm}^{-3}$ if $0\le x \le 30 \,\mu$m and 0 otherwise. The electron density distribution is $n_e(x,y,t_0) = 1.24\, n_0\, \exp{(-y^2/c_e^2)}$ with $c_e = 95 \mu$m. The total number of positrons globally equals that of the electrons but the local net charge imbalance ($c_p\neq c_e$) is purposefully introduced to act as a seed for the instability.The electron-positron cloud is represented by 72 million computational particles (CPs). 
The number density of the background electrons is $n_{\rm pl} = n_0 + 2(n_p-n_e)$ at $t_0$ and their temperature is 50 eV. 

\begin{figure}[!t]
\includegraphics[width=1\columnwidth]{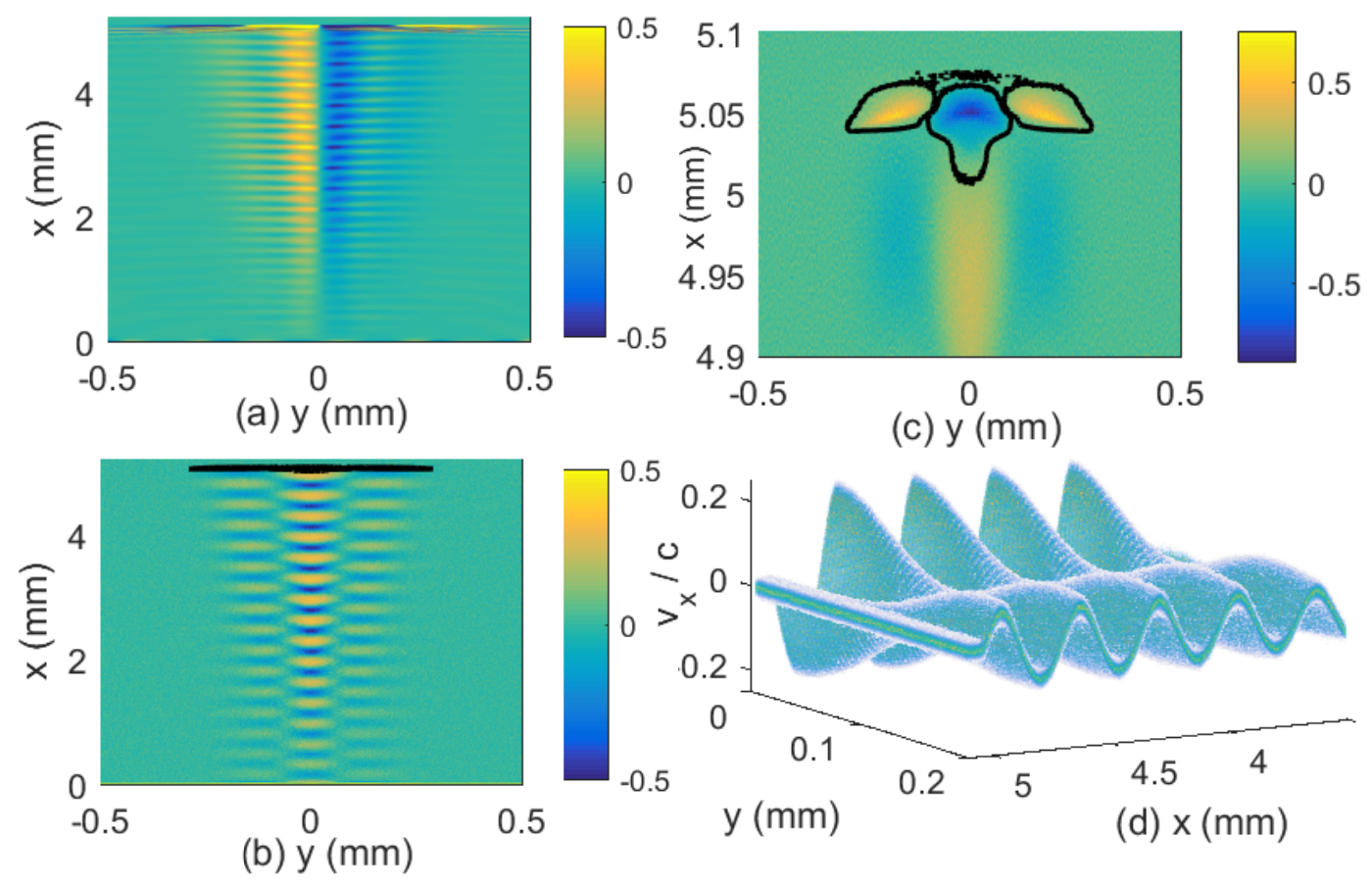}
\caption{Panel (a) shows the amplitude of $B_z(x,y)$ in units of a Tesla. Panel (b) shows the normalized net charge $(n_p - n_b - n_e)/n_0+1$, which takes into account the contribution by the immobile positive background charge. The contour line corresponds to $|n_p-n_e|/n_0=0.01$. Panel (c) is a zoom of (b). The phase space distribution $f_b(x,y,v_x)$ of the background electrons is displayed in (d). All the snapshots are taken at a simulated time of $t=16.9$ ps.}\label{SimFig}
\end{figure}

Figure \ref{SimFig} shows the simulation results at the time 16.9 ps and Ref. \cite{Movie} animates panels (a,b) in time.
The out-of-plane component of the magnetic field (Fig. \ref{SimFig}(a)) grows into two large-amplitude bands, surrounded by two weaker ones. The magnetic field oscillates along x up to a maximum value of $B_0 = 0.6T$ at $x\approx 4.5$ mm.  The cumulative charge density of lepton species in Fig. \ref{SimFig}(b) oscillates in three bands that separate the bands in Fig. \ref{SimFig}(a). Figure \ref{SimFig}(c) demonstrates that the driver of the charge density oscillations is the pair cloud. The peak value of the net charge modulus within the contour exceeds the maximum value of $|n_p(x,y,t_0)-n_e(x,y,t_0)|$ by the factor 3; a filamentation-type instability has spatially separated the cloud's electrons and positrons. 

The cloud's propagation along $x$ transforms the temporal growth of its net charge into the observed spatial growth of $B_z$ and of the charge density perturbations in its wake. The velocity oscillations of the background electrons reach a peak amplitude of 0.25c at $y=0$ in Fig. \ref{SimFig}(d), which is the location where the electrons accumulate in the cloud. The latter have a positive mean velocity and they accelerate the background electrons at $x\approx 5$ mm into the opposite direction. The background electrons are accelerated to increasing values of $x$ by the positrons, which gather in Fig. \ref{SimFig}(c) at $y \approx \pm$ 0.2 mm. 

The moving charged pair cloud induces a return current in the background plasma, which explains the observed strong oscillations of $B_z$ and of the net charge density in the wake of the pair cloud. The only stable charge density wave in an unmagnetized plasma with immobile ions is the Langmuir wave. The large oscillation amplitude together with the two-dimensional structure of the currents have resulted in partially magnetic Langmuir oscillations made of an oscillating and a steady state magnetic component. 

\R{In conclusion, we report on the first experimental observation of the kinetic dynamics of a neutral electron-positron beam. The observed instability results in the generation of strong and long-lived magnetic fields, with an equipartition parameter comparable to what is expected for pair-dominated astrophysical jets. This experimental platform opens the way to accessing fundamental phenomena in basic pair plasma physics and the microphysics of pair-dominated astrophysical scenarios such as magnetic field generation and kinetic dissipation.}

The authors wish to acknowledge support from EPSRC (grants: EP/N022696/1, EP/N027175/1, EP/L013975/1, EP/N002644/1,and EP/P010059/1). The simulations were performed on resources provided by the Swedish National Infrastructure for Computing (SNIC) at HP2CN and on the CINES HPC resource Occigen under the allocation x2016046960 made by DARI/GENCI. 

Data sets are available in [URL to be inserted].


\begin{thebibliography}{100}
\bibitem{comments} V. Tsytovich and C. B. Wharton, Commun. Plasma Phys. Control. Fusion 4, 91 (1978).
\bibitem{Stenson} E. V. Stenson et al., J. Plasma Phys. 83, 595830106 (2017) and references therein.
\bibitem{Blandford} R. D. Blandford and R. L. Znajek, Mon. Not. R. Astron. Soc. 179, 433 (1977).
\bibitem{Begelman} M. C. Begelman, D. Blandford, and J. Rees, Rev. Mod. Phys. 56, 255 (1984).
\bibitem{Goldreich} P. Goldreich and W. H. Julian, Astrophys. J. 157, 869 (1969).
\bibitem{Wardle} J. F. C. Wardle et al., Nature 395, 457 (1998).
\bibitem{Piran} T. Piran, Phys. Rep. 314, 575 (1999).
\bibitem{Racusin} J. L. Racusin et al., Nature 455, 183 (2008).
\bibitem{Bret} A. Bret et al., Phys. Plasmas 17, 120501 (2010).
\bibitem{Medvedev} M. V. Medvedev and A. Loeb, Astrophy. J. 526, 697 (1999).
\bibitem{Reville} B. Reville et al. Plasma Phys. Contr. F. 48, 1741 (2006).
\bibitem{Lyubarsky} Y. Lyubarsky and D. Eichler,  Astrophys. J. 647, 1250 (2006). 
\bibitem{Gruzinov} A. Gruzinov Astrophys. J. 563, 15 (2001).
\bibitem{Ferriere} K. M. Ferriere, Rev. Mod. Phys. 73, 1031 (2001).
\bibitem{Galama} T. J. Galama et al., Nature 398, 394  (1999).
\bibitem{Vreeswijk} P. M. Vreeswijk et al., Astrophys. J. 523, 171 (1999).
\bibitem{Waxman} E. Waxman, Astrophys. J. 485, L5 (1997).
\bibitem{Wijers} R. A. M. J. Wijers and T. J. Galama, Astrophys. J. 523, 177 (1999).
\bibitem{Sari} R. Sari et al., Astrophys. J. 473, 204 (1996).
\bibitem{Meszaros} P. Meszaros et al., Astrophys. J. 415, 181 (1993).
\bibitem{Chang} P. Chang et al., Astrophys. J. 674, 378 (2008).
\bibitem{Muggli}P. Muggli et al. ArXiv: 1306.4380v1 (2013).
\bibitem{Silva} L. Silva et al. Astrophys. J. 596, L121 (2003).
\bibitem{Dieckmann} M. E. Dieckmann et al. Astronomy \& Astrophysics 577, A137 (2015).
\bibitem{Chen} H. Chen et al., Phys. Rev. Lett. 114, 215001 (2015).
\bibitem{Pedersen} T. S. Pedersen et al. New J. Phys. 4, 035010 (2012).
\bibitem{Liang} E. Liang et al. Sci. Reports 5, 13968 (2015).
\bibitem{SarriNcomm} G. Sarri et al., Nat. Comm. 6, 6747 (2015).
\bibitem{SarriPPCF} G. Sarri et al., Plasma Phys. Contr. F. 55, 124017 (2013).
\bibitem{SarriPRL} G. Sarri et al. Phys. Rev. Lett. 110, 255002 (2013).
\bibitem{SarriNJP} G. Sarri et al. New J. Phys.12, 045006 (2010). 
\bibitem{Borghesi} M. Borghesi et al. Phys. Plasmas 9, 2214 (2002).
\bibitem{Gemini} C. J. Hooker et al., J. Phys. IV 133, 673 (2006).
\bibitem{Esarey} E. Esarey et al., Rev. Mod. Phys. 81, 1229 (2009).
\bibitem{Clayton} C. E. Clayton et al., Phys. Rev. Lett. 105, 105003 (2010).
\bibitem{Kneip} S. Kneip et al., Phys. Rev. Lett. 103, 035002 (2009). 
\bibitem{SarriPPCF2} G. Sarri et al., Plasma Phys. Contr. F. 59, 014015 (2017). 
\bibitem{BorghesiRMP} A. Macchi et al., Rev. Mod. Phys. 85, 751 (2013).
\bibitem{suppTIME} See Supplemental Material at [URL will be inserted by publisher] for the time-resolved evolution of the phenomenon for each percentage of positrons in the beam and the typical spatial distribution of the unperturbed proton beam.
\bibitem{SRIM} http://www.srim.org/
\bibitem{FLUKA} G. Battistoni et al. AIP conf. proc. 896, 31 (2007).
\bibitem{HYADES} http://casinc.com/hyades.html
\bibitem{suppHYADES} See Supplemental Material at [URL will be inserted by publisher] for a summary of the results of the hydrodynamic simulation on photo-ionisation of the background plasma.
\bibitem{Joshi} C. Joshi et al., Phys. Plasmas 9, 1845 (2002).
\bibitem{Smyth} A. Smyth et al., Phys. Plasmas 23, 063121 (2016).
\bibitem{filamentation} K. M. Watson et al., Phys. Fluids 3, 741 (1960). 
\bibitem{Sironi} L. Sironi et al., Astrophys. J. 787, 49 (2014).
\bibitem{suppCHARGE} See Supplemental Material at [URL will be inserted by publisher] for the charge and number density of the EPB after propagation through the background gas, as obtained from matching 3-Dimensional Particle-In-Cell simulations.
\bibitem{Arber15} T. D. Arber et al., Plasma Phys. Controll. Fusion \textbf{57}, 113001 (2015).
\bibitem{Movie} See Supplemental Material at [URL will be inserted by publisher] for a time-animation of panels (a) and (b) of Fig. \ref{SimFig} for the times 0.7 ps $< t <$ 33.6 ps. 

\end{thebibliography}
\end{document}